
\input phyzzx
\baselineskip=18pt
\hfill {CU-TP-647}\break
\FRONTPAGE
\strut\hfill {SNUTP94-71} \break 
\strut\hfill { hep-th/9408079} \vglue 0.8in
\font\twelvebf=cmbx12 scaled\magstep2
\font\smallrm=cmr5
\vskip 0.0in
\centerline {\twelvebf The Chern-Simons
Coefficient in the Higgs Phase}

\vskip .5in
\centerline{{\it Hsien-Chung Kao} {\rm and} {\it  Kimyeong Lee }}
\vskip .1in
\centerline{ Physics Department, Columbia University}
\centerline{ New York, N.Y.  10027, U.S.A.}
\vskip 0.3in
\centerline{ \it Choonkyu Lee}
\vskip .1in
\centerline{ Physics Department and Center for Theoretical Physics}
\centerline{ Seoul National University,  Seoul, 151-742, Korea}
\vskip 0.3in
\centerline{\it Taejin Lee}
\vskip 0.1in
\centerline{ Physics Department, Kangwon National University}
\centerline{ Chuncheon, 200-701, Korea}
\vskip 0.4in
\centerline{\bf ABSTRACT}
\vskip 0.1in

We study one-loop corrections to the Chern-Simons coefficient $\kappa$
in abelian self-dual Chern-Simons Higgs systems and their $N=2$ and
$N=3$ supersymmetric generalizations in both symmetric and asymmetric
phases.  One-loop corrections to the Chern-Simons coefficient of these
systems turn out to be integer multiples of $1/4\pi$ in both phases.
Especially in the maximally supersymmetric $N=3$ case, the correction
in symmetric phase vanishes and that in asymmetric phase is
$\kappa/(2\pi |\kappa|)$.  Our results suggest that nonabelian
self-dual systems might enjoy  similar features.  We also discuss
various issues arising from our results.

\vfill \endpage


\def\pr#1#2#3{Phys.  Rev.  {\bf D#1}, #2 (19#3)}
\def\prl#1#2#3{Phys. Rev.  Lett.  {\bf #1}, #2 (19#3)}

\def\np#1#2#3{Nucl.  Phys.  {\bf B#1}, #2 (19#3)}
\def\pl#1#2#3{Phys.  Lett.  {\bf B#1}, #2 (19#3)}
\def\ibid#1#2#3{ {\it ibid.} {\bf #1}, #2 (19#3)}
\def\tr{\,{\rm tr}}
\def\half{ {\smallrm {1\over 2} }}
\def\vh{\hat{\varphi}}
\def\vp{\varphi}



\REF\rColeman{S.  Coleman and B.  Hill, \pl{159}{184}{85}.}

\REF\rSemenoff{G.W.  Semenoff, P.  Sodano and Y.-S.  Wu,
\prl{62}{715}{89}; W.  Chen, \pl{251}{415}{91}; V.P.  Spiridonov and
F.V.  Tkachov, \ibid{260}{109}{91}; D.K. Hong, T. Lee and S.H. Park,
\pr{48}{3918}{93}.  } 

\REF\rKhleb{S.Y.  Khlebnikov, JETP letters {\bf 51}, 81, (1990); V.P.
Spiridonov.  \pl{247}{337}{90}.  } 

\REF\rDeser{S.  Deser, R.  Jackiw and S.  Templeton, Ann.  Phys.
(N.Y.)  {\bf 140}, 372 (1982).}

\REF\rPisarski{R.D.  Pisarski and S.  Rao, \pr{32}{2081}{85}; G.
Giavirini, C.P.  Martin and F.  Ruiz Ruiz, \np{381}{222}{92}.}

\REF\rKhlebn{ S.Y.  Khlebnikov and M.E.  Shaposhnikov,
\pl{254}{148}{91}; A.  Khare, R.B.  MacKenzie, P.K.  Panigrahi and
M.B.  Parajape, ``Anomalies via spontaneous symmetry breaking in 2+1
dimensions,'' Univ.  de Montr\'eal Preprint, UdeM-LPS-TH-93-150,
hep-th/9306027}

\REF\rHong{ J.  Hong, Y.  Kim and P.Y.  Pac, \prl{64}{2230}{90}; R.
Jackiw and E.J.  Weinberg, \ibid{64}{2234}{90}; R.  Jackiw, K.  Lee
and E.J.  Weinberg, \pr{42}{3488}{90}; Y.  Kim and K.  Lee,
\ibid{49}{2041}{94}.}

\REF\rClee{C.  Lee, K.  Lee, and E.J.  Weinberg, \pl{243}{105}{90};
E.A.  Ivanov,\ibid{268}{203}{91}; S.J.  Gates and H.  Nishino,
\ibid{281}{72}{92}.}

\REF\rKao{H.-C.  Kao and K.  Lee, \pr{46}{4691}{92}; H.-C.
Kao,``Self-Dual Yang-Mills Chern-Simons Higgs Systems with an N=3
Extended Supersymmetry,'' Columbia preprint, CU-TP-595 (1993), to
appear in Phys.  Rev.  D.}

\REF\rKlee{ K.  Lee, \prl{66}{553}{91}; K.  Lee, \pl{255}{381}{91};
G.V.  Dunne, \ibid{324}{359}{94}; H.-C.  Kao and K.  Lee, ``Self-Dual
SU(3) Chern-Simons Higgs Systems,'' Columbia preprint, CU-TP-635
(1994); G.  Dunne, ``Vacuum Mass Spectra for SU(N) Self-Dual
Chern-Simons-Higgs Systems,'' Univ.  Connecticut preprint UCONN-94-4
(1994).}

\REF\rIpeko{ Y.  {\.  I}peko{\v g}lu, M.  Leblanc and M.T.  Thomaz,
Ann.  Phys. (N.Y.), {\bf 214}, 160 (1992).}

We consider one-loop correctons to the Chern-Simons coefficient in
abelian Chern-Simons Higgs systems.  In broken phase of these systems
there are topological vortices of fractional spin and statistics.  The
quantum correction to the Chern-Simons coefficient could lead to a
change in vortex dynamics given in the tree approximation.  If there
are charged particles in broken phase, there would be a nontrivial
Aharonov-Bohm-type interaction between vortices and charged particles,
which could also get a quantum correction.  Thus, to find the quantum
correction to the Chern-Simons coefficient would be  interesting
as a step to understand quantum physics in broken phase.

There have been recently some controversies related to the nature of
the correction to the Chern-Simons coefficient.  In symmetric phase of
abelian systems, there is a theorem that the correction originates
from the fermion one-loop and no more [1].  When there are massless
charged particles or when the gauge symmetry is spontaneously broken,
the theorem breaks down and the loop corrections turn out to be in
general complicated functions of couplings and particle masses [2,3].
When the gauge symmetry is nonabelian, the coefficient should be
quantized to have a consistent quantum theory[4].  While this is
indeed the case in the symmetric phase [5], the quantization is not
respected by one-loop correction in asymmetric phase [6].  However,
there has been no clear understanding of this puzzle.

In this paper, we calculate one-loop correction to the Chern-Simons
coefficient in abelian self-dual Chern-Simons Higgs systems and their
$N=2,N=3$ supersymmetric generalizations [7,8,9] in both symmetric and
asymmetric phases.  The corrections in these systems turn out to be
`quantized' in the sense that they are integer multiples of the Dirac
fermion correction, $1/4\pi $.  This suggests immediately an
interesting possibility that the one-loop corrections in nonabelian
self-dual systems [10] and their supersymmetric systems [8,9] would be
also quantized.  While the quantum consistency condition of nonabelian
systems does not imply the quantization of the Chern-Simons
coefficient in asymmtric phase, we will argue that our result is no
coincidence and that there may be other reasons why the correction
should be quantized.  These points and other aspects arising
from our calculation would be discussed at the end.

We first start with a simple model of a gauge field $A_\mu$, a complex
Higgs field $\phi$, and a complex fermion field $\psi $.  The most
general symmetric lagrangian which is renormalizable is given as
$$ \eqalign{ {\cal L}
= &\ - {1\over 4e^2} F_{\mu\nu}^2 + {\kappa \over 2}
\epsilon^{\mu\nu\rho}A_\mu \partial_\nu A_\rho + |D_\mu \phi|^2 +
i\bar{\psi}\gamma^\mu D_\mu \psi \cr &\ -U(|\phi|) - (M + 2 g_1 |\phi|^2 )
\bar{\psi}\psi - g_2 [ \phi^2 \bar{\psi}\psi^* + \phi^{*2}
\bar{\psi}^*\psi] \cr}
\eqno\eq $$
where $D_\mu = \partial_\mu + iA_\mu$ and all coupling constants are
real.  (An additional possible term $g' i[\phi^2\bar{\psi}\psi^* -
\phi^{*2}\bar{\psi}^*\psi]$ can be transformed to the last term in
Eq.(1) by a suitable global phase rotation on $\psi$.)  For this
theory to be renormalizable, a simple dimensional counting shows that
$U(|\phi|)$ is at most sixth order in $\phi$.  The metric is given by
$\eta_{\mu\nu} = {\rm diag}(1,-1,-1,)$ and $\epsilon^{012} =
\epsilon_{012} = 1$.  The gamma matrices are pure imaginary,
$\gamma^\mu= (\tau^2, i\tau^3, i\tau^1 )$, and satisfy $
\gamma^\mu \gamma^\nu = \eta^{\mu\nu} -i\epsilon^{\mu\nu\rho}
\gamma_\rho$.

We are interested in calculating one-loop correction to the photon
propagator in a constant background scalar field.  We rewrite the
scalar and spinor fields in real and Majorana fields, $\phi = (\phi_R +
i \phi_I)/\sqrt{2}$ and $\psi = (\psi_R + i\psi_I)/\sqrt{2}$.  We
separate the scalar fields $\phi_a$ into the constant background field
$\vp_a$ and the quantum fluctuations $\phi_a$.  The background lagrangian is
then given by
$$ {\cal L}_B = {\cal L}(A_\mu, \vp_a + \phi_a, \psi)
-{\cal L}(0, \varphi_a, 0 ) - \phi_a {\partial {\cal L} \over \partial
\phi_a } (0,\vp_a,0)
\eqno\eq $$
In addition, we use the $R_\xi$
gauge fixing,
$$ {\cal L}_{gf} = -{1\over 2\xi } (\partial^\mu A_\mu
+ \xi \epsilon_{ab} \phi_a \vp_b) )^2
\eqno\eq$$
which leads to the Fadeev-Popov ghost lagrangian
$$ {\cal L}_{FP}
= \bar{\eta} (-\partial_\mu^2 - \xi (\vp_a^2 + \vp_a \phi_a))\eta
\eqno\eq $$
The ghost loop contributes to the photon propagator in
one-loop via the $\phi_a$ field tadpole diagram, but no
correction to the Chern-Simons coefficient.

Putting the lagrangian (2) and the gauge fixing terms (3) and (4)
together, we get the quadratic term
$$ \eqalign{ {\cal L}_0 &\ = \half A^\mu\left(
( {1\over e^2} \partial_\rho^2 + \vp^2_a) \eta_{\mu\nu} - ({1\over
e^2} - {1\over \xi}) \partial_\mu \partial_\nu - \kappa
\epsilon_{\mu\nu\rho} \partial^\rho \right) A^\nu \cr
&\ + \half
\phi_a \left( -\partial_\mu^2 \delta_{ab} - m_1^2 \vh_a\vh_b - (m_2^2
+ \xi \vp_c^2) (\delta_{ab} - \vh_a \vh_b) \right) \phi_b \cr
&\ +
\half \bar{\psi}_a \left( i \gamma^\mu \partial_\mu \delta_{ab} - M_1
\vh_a\vh_b -M_2 (\delta_{ab}- \vh_a \vh_b)\right) \psi_b \cr
&\ +
\bar{\eta} (-\partial_\mu^2 - \xi \vh_a^2)\eta\cr}
\eqno\eq $$
where the terms quadratic in $\phi_a$ of $U(\sqrt{(\vp_a+ \phi_a)^2})$
is $ m_1^2 \vh_a\vh_b \phi_a \phi_b/2 + m_2^2 (\delta_{ab}
-\vh_a\vh_b)\phi_a\phi_b /2$, and we have denoted $M_1 \equiv M + g_1
\vh_a^2$, and $M_2 \equiv g_2 \vh_a^2$.  The interaction parts
involving the gauge field become
$$ \eqalign{ {\cal L}_I = &\ \vp_a \phi_a A_\mu^2 -
\half \phi_a^2 A_\mu^2 - { i \over 2} A_\mu \epsilon_{ab} \bar{\psi}_a
\gamma^\mu \psi_b \cr &\ - \epsilon_{ab}\phi_b \partial^\mu \phi_a
A_\mu - \xi  \vp_a\phi_a \bar{\eta}\eta \cr}
\eqno\eq $$
{}From the quadratic terms (6) it is straightforward to get the
propagators for the gauge, scalar and fermion fields,
$\Delta_{\mu\nu}(p) , D_{ab}(p), S_{ab}(p)$.  Here we just display the
propagator for the gauge field,
$$ \eqalign{ \Delta_{\mu\nu}(p) &\ = {-i (p^2/e^2 - \vp_a^2) \over
(p^2/e^2 - \vp_a^2)^2 - \kappa^2 p^2} \left(\eta_{\mu\nu} - {p_\mu
p_\nu \over p^2} \right) \cr
&\ + {\kappa \epsilon_{\mu\nu\rho} p^\rho
\over (p^2/e^2- \vp_a^2)^2 - \kappa^2p^2} + {-i \xi p_\mu p_\nu \over
p^2 (p^2/e^2 - \xi \vp_a^2)} \cr}
\eqno\eq $$

By the Lorentz invariance and gauge invariance, the full inverse photon
propagator may be expressed as
$$ \eqalign{ i\Delta_{\mu\nu}^{-1}(p) = &\
(p^2g_{\mu\nu} - p_\mu p_\nu) \Pi_1(p^2) + i
\epsilon_{\mu\nu\rho}p^\rho \Pi_2(p^2) \cr
&\ + p_\mu p_\nu
\Pi_3(p)\cr}
\eqno\eq $$
The one-loop graphs would be linearly divergent, but become finite
after the Pauli-Villar regularization.  With the convention $D_\mu =
\partial_\mu + iA_\mu$, there is no wave function renormalization of
the gauge field.  Among many Feynman graphs contributing to the photon
self-energy, there are two graphs for possible corrections to the
Chern-Simons term $\Pi_2$: a mixed loop of the $\phi$ and gauge fields
and a fermion loop.  The Chern-Simons coefficient appears in the full
photon propagators as
$$ \Pi_2(p^2) = \kappa + \Pi^B(p^2) +
\Pi^F(p^2)
\eqno\eq $$
Here we have chosen our regularization so that $\Pi_2(p^2 \rightarrow
-\infty) = \kappa $; at short distance, the correction would vanish.
The one-loop corrected Chern-Simons coefficient would be then
$\Pi_2(p=0)$.

If we try to understand the Chern-Simons coefficient in an effective
action, there would be many gauge invariant terms in the effective
action which could appear as a Chern-Simons term at large distance in
{\it asymmetric} phase.  One of such terms is $\epsilon^{\mu\nu\rho}
i(D_\mu\phi^* \phi - \phi^*D_\mu \phi) F_{\nu\rho}$ with an
appropriate coupling constant.  Similar term for fermions describes
the interaction between an anomalous magnetic moment and the magnetic
field.  In asymmetric phase, this term would contribute to the
Chern-Simons coefficient.  The bosonic correction to the Chern-Simons
coefficient can be seen as the result of many such gauge-invariant
terms in the effective action.  Since these terms are invariant also
under large gauge transformations, there seems no need for such
bosonic corrections to the Chern-Simons term should be quantized in
general.  (Similar point of view appeared in the first paper of
Ref.[6].)  However, we will later argue that there may be other
reasons why the sum of these bosonic corrections should be quantized.

{}From the interaction terms (6), we get the mixed bosonic 1-loop
contribution to the photon self-energy
$$ i\pi_{\mu\nu}^B(p) = 4i^2 \vp_a \vp_b \int {d^3q
\over (2\pi)^3} \Delta_{\mu\nu} (q) D_{ba}(q-p)
\eqno\eq $$
In pure
Chern-Simons limit ($e^2\rightarrow \infty $) the above bosonic
correction leads to
$$ \Pi^B(p^2) = {\mu \over 4\pi} \int_0^1
d\alpha {1 \over \sqrt{ (1-\alpha) \mu^2 + \alpha m_1^2 -
\alpha(1-\alpha) p^2 } }
\eqno\eq $$
where $\mu \equiv
\vp_a^2/\kappa$.  Taking the zero momentum limit, we get the bosonic
correction to the Chern-Simons coefficient,
$$ \Pi^B(0) = {1\over
4\pi} \left\{ {4\mu ( 2|\mu| +|m_1| ) \over 3 ( |\mu|+|m_1|)^2 }
\right\}
\eqno\eq $$
In the limit where $\kappa \rightarrow 0$ or
$\mu \rightarrow \infty$, it becomes $2\kappa / (3|\kappa|) \times 1/4\pi$,
which is an old result [3].  When two masses $\mu, m_1$ become
identical, $\Pi^B(p^2=0) = \kappa/( 4\pi |\kappa|)$.
Such coincidence of the two masses is realized in  abelian self-dual
Chern-Simons Higgs systems[7] and so we have just  shown that these
self-dual systems in asymmetric phase would acquire a quantized
one-loop correction to the Chern-Simons coefficient.

{}From the interaction terms (6), it is also straightforward to get the
fermionic one-loop contribution
$$ i \pi^F_{\mu\nu}(p) = -
\epsilon_{ab} \epsilon_{cd} \int {d^3q \over (2\pi)^3} \tr \gamma_\mu
S_{bc}(q) \gamma_\nu S_{da}(q-p)
\eqno\eq $$
which leads to the following correction to $\Pi_2$:
$$ \Pi^F(p^2) = -{1 \over 4\pi} \int_0^1
d\alpha {M_1 \alpha + M_2(1-\alpha) \over \sqrt{M_1^2\alpha
+M_2^2(1-\alpha) - p^2 \alpha(1-\alpha) } }
\eqno\eq $$
Taking the zero momentum limit, we see that
the fermionic one-loop correction to the coefficient is given as
$$ \Pi^F(0) = - {1 \over 6\pi} \left\{ { |M_1|(M_1 + 2M_2) +
|M_2|(2M_1+ M_2) \over (|M_1| + |M_2|)^2} \right\}
\eqno\eq $$
We will later be interested in two special cases.  When $M_1=M_2$,
we have a Dirac fermion of mass $M_1$ and the correction
becomes $ - M_1/ (4\pi |M_1|)$.  When $M_1=-M_2$ on the other hand,
our fermion system consists of two Majorana fermions of spin $1/2$ and $-1/2$
and then the correction (15) vanishes.

Let us now apply the above results to an $N=3$ self-dual Chern-Simons
Higgs system[9].  Once we know the correction in the $N=3$ theory,
we will see that it is trivial to read off the correction for systems with
lesser supersymmetry.  The lagrangian for the $N=3$ theory is given
by
$$ \eqalign{ {\cal L} &\ = {\kappa \over 2} \epsilon^{\mu\nu\rho}
A_\mu \partial_\nu A_\rho + |D_\mu\phi_a|^2 + i \bar{\psi} \gamma\cdot
D \psi + i\bar{\chi}\gamma\cdot D \chi \cr
&\ - {1\over \kappa^2}
|\phi_a|^2[(|\phi_a|^2)^2 + 2v^2(|\phi_1|^2-|\phi_2|^2)+ v^4] \cr
&\ -
{v^2\over \kappa} (\bar{\psi}\psi -\bar{\chi}\chi) + {3|\phi_a|^2\over
\kappa} (\bar{\psi}\psi +\bar{\chi}\chi) \cr
&\ -{4\over \kappa}
(\phi_1\bar{\psi}-\phi_2\bar{\chi})(\phi_1^*\psi-\phi_2^*\chi) \cr
&\
-{1\over\kappa}(\phi_1\bar{\psi}-\phi_2\bar{\chi})
(\phi_1\psi^*-\phi_2\chi^*) \cr
&\
-{1\over\kappa}(\phi_1^*\bar{\psi}^*-\phi_2^*\bar{\chi}^*)
(\phi_1^*\psi-\phi_2^*\chi) \cr}
\eqno\eq $$
where $D_\mu = \partial_\mu + iA_\mu$.  This lagrangian is invariant
under an abelian gauge symmetry and an additional global symmetry
corresponding to a uniform phase rotation of
$\phi_1,\phi_2^*,\psi,\chi^*$.  The central charge of the
supersymmetric algebra is proportional to the magnetic flux.  This
theory is obtained by maximally supersymmetrizing the abelian
self-dual Chern-Simons Higgs systems.  There are degenerate symmetric
and asymmetric vacua, and the self-dual configurations are topological
vortices in asymmetric phase and nontopological q-balls in symmetric
phase [7].

In symmetric phase, there are no particle degrees of freedom for the
gauge field.  Two fermions are of the opposite mass $\pm v^2/\kappa$
and also of the opposite spin. Scalar bosons carry mass $v^2/\kappa$.
The matter fields form a reduced supermultiplet $(2,4,2)$ of spin
$(1/2,0,-1/2)$, saturating the energy bound given by the central
charge.  Explicit one-loop calculation is rather straightforward.  In
addition, our theory is finite at one-loop and needs no
regularization.  Since the two Dirac fermions $\psi,\chi$ come with
the opposite mass, there is no correction to the Chern-Simons term.
There is a nontrivial generation of the Maxwell term,
$$ i\pi_{\mu\nu} = {i\over \pi p^2} \left\{
{|m| \over 2} + {m^2 \over 4 p} \ln \left( {|m| -p \over |m| +
p} \right) \right\} ( p^2 \eta_{\mu\nu} - p_\mu p_\nu )
\eqno\eq $$
where $m\equiv 2v^2/\kappa$.  Naively, this term leads to a Maxwell
term when $p\rightarrow 0 $, leading to a dynamical generation of
photons.  However, the photon mass is of order $|m| \kappa$ which lies
far above the branch point $p= |m|$ for the small coupling constant
$\kappa>>1$.  Hence, there is no dynamical generation of photons at
least in the perturbative regime.  Calculating one-loop correction to
mass of various fields is straightforward and follows the standard
procedure.  In symmetric phase, all one-loop corrections to masses of
elementary particles turn out to vanish.

In  asymmetric phase, the calculation is more delicate. Introducing the
new variable $\phi_2 = v + (f + ig)/\sqrt{2}$, the free lagrangian in
the broken phase becomes
$$ \eqalign{ {\cal L}_1 = &\ {\kappa \over
2} \epsilon^{\mu\nu\rho} A_\mu \partial_\nu A_\rho + |\partial_\mu
\phi_1|^2 + \half (\partial_\mu f)^2+\half (\partial_\mu g)^2 \cr
&\ +
i\bar{\psi} \gamma\cdot \partial_\mu \psi + {i \over 2} [ \bar{\chi}_R
\gamma\cdot \chi_R + \bar{\chi}_I \gamma\cdot \partial \chi_I ] \cr
&\
+ {\kappa m \over 2} A_\mu^2 - m^2 |\phi_1|^2 - \half m^2 f^2 +
m\bar{\psi}\psi \cr &\ - {m \over 2} (\bar{\chi}_R\chi_R -
\bar{\chi}_I\chi_I) + \sqrt{2} v \partial^\mu g A_\mu \cr}
\eqno\eq $$
We need a gauge fixing term (3) and the corresponding ghost lagrangian
(4) with $\vp_a = \sqrt{2} v \delta_{a0}$.  The gauge field absorbs
the Goldstone boson to become a massive vector boson.  The particles
form a supermultiplet $(1,3,3,1)$ of spin $(1,1/2,0,-1/2)$.

Now it is straightforward to calculate the one-loop correction to the
Chern-Simons coefficient.  We may in fact use our earlier results (12)
and (15).  First we note that there is no correction from $\chi_{R,I}$
because their masses are opposite to each other in the sign.  The
gauge boson and the Higgs field have the same mass, i.e., $\mu=m,
m_1=m$, and so they yield the correction, $\kappa/(4\pi |\kappa|)$.
For the $\psi$ fermions, the parameters $M_{1,2}$ becomes $-m$ and its
contribution to the Chern-Simons coefficient is $m/(4\pi|m|)$.  Thus,
the bosonic and fermionic contributions come with the {\it same} sign.
One loop corrected Chern-Simons coefficient (9) is then
$$ \Pi_2( p=0) = \kappa + {1\over 2\pi} {\kappa \over
|\kappa|}
\eqno\eq $$
This is the main result of this letter.  Note that the correction is
quantized and has the same sign as the classical term.  Any higher
order correction, if exists, would be of order $1/\kappa$ and spoil
the quantization feature in the small coupling limit $\kappa>>1$.  If
there are many coupling constants (as in non-supersymmetric or
non-self-dual theories), the loop expansion would be more complicated.

The $N=2$ supersymmetric systems [8] can be obtained from Eq.(16) by
dropping the terms depending on $\phi_1, \chi$.  Thus, there would be
a nonzero correction $-\kappa/(4\pi |\kappa|) $ in the symmetric phase
and the correction (19) in the asymmetric phase.

Since our result depends crucially on the sign of the terms
in the lagrangian (16) and the sign of the results (12) and
(15), we have performed an independent check to confirm the
result (19).  As the Chern-Simons coefficient gets a nonzero
one-loop correction (19) in asymmetric phase, there would be
also a correction to the mass of massive vector bosons.  If
the supersymmetry is preserved to one-loop, the mass
corrections to other fields would be identical to that of
massive gauge bosons.  This turns out to be indeed the case.

 Since the calculation is rather straightforward and the method is in
standard textbooks, we simply summerize our findings.  First, we
calculated the correction to the vacuum expectation value both by the
Feynman diagram method and by the effective potential method and
obtained the identical result, $ \vev{f} = -\sqrt{2} (m-\sqrt{2\xi
v^2} )/(16\pi v) $.  This correction is again finite because of the
$N=3$ supersymmetry.  (For the case with $N=2$, it is infinite and
needs a renormalization [11].)  The vacuum energy degeneracy between
symmetric and asymmetric phases is preserved in one-loop.  Second, we
calculated the correction to the propagators of various fields in
asymmetric phase.  Mass of an elementary particle would appear as the
pole of each propagator and is gauge invariant and independent of the
gauge fixing parameter $\xi$.  The corrections to the propagators
assume particularly simple forms for the choice $2\xi v^2=m^2$.  For
example, the photon propagator correction in asymmetric phase would be
$$ \pi_{\mu\nu} = \left\{ - \eta_{\mu\nu} p^2 + p_\mu p_\nu
+ 2 im \epsilon_{\mu\nu\rho} p^\rho \right\} {1\over 4\pi p} \ln {2|m|
+ p \over 2|m| - p}
\eqno\eq $$
{}From Eq.(20) it is easy to confirm the result (19).  We can also
calculate the mass shift from the above expression (20).  The mass of
the vector boson gets shifted from $m$ to $m+\Delta m$ with $ \Delta m
= - 3m (\ln 3) /(4\pi \kappa) $.  We have found the identical mass
shift for other particles in asymmetric phase.

In this letter we have calculated one-loop correction to the
Chern-Simons coefficient in self-dual Chern-Simons Higgs systems and
their $N=2,N=3$ supersymmetric generalizations in symmetric and
asymmetric phases.  These corrections turn out to be ``quantized'' in
the sense that they are integer times the Dirac fermion contribution
$1/4\pi$.  While it is not clear whether this feature will be
preserved in higher loops, our result suggests various interesting
possibilities.  Some self-dual models might have higher order
correction of order $1/\kappa$ which might spoil this `quantization'
feature.  Some supersymmetric models, especially the $N=3$ case, might
have no higher order corrections.  To support or refute these
speculations need a further effort.  The Maxwell kinetic term for the
gauge field complicates our result and we hope to explore this theory
in near future.

Our results suggest that one-loop correction to the Chern-Simons
coefficient in nonabelian self-dual Chern-Simons Higgs systems or
their supersymmetric generalizations might be quantized.  Again, there
may be no higher loop correction in some of these nonabelian models.
If we break supersymmetry or self-duality, the one-loop correction
would not in general be quantized.  There seem to be two possibilities
in such cases.  The first possibility is that the corrected
Chern-Simons coefficient might again be quantized if we include all
higher loop corrections and possible nonperturbative corrections. The
second possibility is that the fully corrected Chern-Simons
coefficient is not quantized, with a possible theoretical difficulty
discussed below.

As argued before, the bososnic contribution to the Chern-Simons
coefficient originates from the parity violating gauge-invariant terms
in the effective action.  Hence the bosonic correction does not need
to be quantized to be consistent with the global gauge
transformations.  On the other hand, there may be another reason why
one demands the correction to the Chern-Simons coefficient to be
quantized.  First, we can consider an abelian system.  If the
Chern-Simons coefficient is a rational number, we can put vortices on
a large sphere, being consistent with quantized magnetic flux, charge
and angular momentum.  However, a nonquantized correction to the
Chern-Simons coefficient could make the system quantum mechanically
inconsistent one. Second, in nonabelian systems the flux and charge
quantization is necessary to get a finite representation of a braid
group. Such feature would be spoiled if the quantum correction is not
quantized appropriately.

Finally there are some interesting implications we like to point out.
First, the one-loop correction in asymmetric phase is nonzero and
quantized.  This would lead to the corrections to the spin-statistics
of vortices and to the interaction between vortices and charged
particles in broken phase.  Second, there is a diffecence between
corrections in symmetric and asymmetric phases.  If we break the
degeneracy between symmetric and asymmetric vacua or if we think about
the theory on a large sphere, there could be tunneling between two
vacua.  The possible tunneling does not seem consistent naively with
the fact that the corrected Chern-Simons coefficient is different
between two phases.

\vskip 0.2in

\centerline{\bf Acknowledgement }

We thank A.  Mueller and A.  Polychronakos for simulating
discussions.  The work by H.-C.K.  and K.L.  is supported in
part by U.S.  DOE, the NSF Presidential Young Investigator
program and the Alfred P.  Sloan Foundation.  The work by
C.L.  and T.L.  is supported by Korea Science and
Engineering Foundation (through the Center for Theoretical
Physics, SNU) and the Basic Science Research Instititute
Program, Ministry of Education, Korea (Project No.
BSRI-94-2418 and No.  BSRI-94-2401).

\vfill \endpage

\refout

\end